\documentclass[%
reprint,
twocolumn,
showpacs,preprintnumbers,
amsmath,amssymb,
aps,pre, 
nofootinbib,
longbibliography, 
floatfix,
]{revtex4-1}

\usepackage{enumitem}  
\usepackage{graphicx}
\usepackage{epstopdf}
\usepackage{dcolumn}
\usepackage{bm}
\usepackage{hyperref}
\usepackage{color,soul}
\usepackage{lipsum}  
\usepackage[export]{adjustbox}

\newcommand{\ie}{\textit{i.e.,}\ }
\newcommand{\eg}{\textit{e.g.}\ }
\newcommand{\md}{\mathrm{d}}
\newcommand{\etal}{\textit{et al.}}

\emergencystretch=\maxdimen
\begin{document}


\title{Laser frequency upconversion in plasmas with finite ionization rates }
\author{Kenan Qu}
\author{Nathaniel J. Fisch}
\affiliation{%
	Department of Astrophysical Sciences, Princeton University,  Princeton, New Jersey 08544, USA
}%

\date{\today}

\begin{abstract}
	Laser frequency can be upconverted in a plasma undergoing ionization. For finite ionization rates, the laser pulse energy is partitioned into a pair of counter-propagating waves and static transverse currents. The wave amplitudes are determined by the ionization rates and the input pulse duration. The strongest output waves can be obtained when the plasma is fully ionized in a time that is shorter than the pulse duration. The static transverse current can induce a static magnetic field with instant ionization, but it dissipates as heat if the ionization time is longer than a few laser periods. 
	This picture comports with experimental data, providing a description of both laser frequency upconverters as well as other laser-plasma interaction with evolving plasma densities. 
\end{abstract}

\maketitle

\section{Introduction}

Traditional methods of converting the electromagnetic wave frequency using nonlinear optical crystals are usually restricted by requiring phase matching and polarized light~\cite{Boyd_1968}. They are not available for all light sources. The conversion processes are efficient only at sufficiently high optical intensities, but not so high as to cause thermal damage to the nonlinear crystals. Another possibility to convert laser frequencies is using temporally dynamic media~\cite{morgenthaler1958velocity, wilks1988frequency, mendoncca2000book, nerukh2012non, kalluri2016electromagnetics, Maslov_2018}. It does not need phase matching and it works with lasers not suitable for the traditional nonlinear optical methods. If the medium is ionizing plasma, the method can also sustain high intensities above the ionization threshold limit. ``Proof-of-principle'' numerical simulations predicted~\cite{Manuscript_Chirped} that this method may create ``tabletop'' sources of coherent x rays out of near-infrared lasers at high efficiency and low cost. However, the existing theories only describe either ``flash ionization''~\cite{wilks1988frequency, Kenan_2018_upshift}, with which the plasma is fully ionized instantaneously, or ``adiabatically gradual ionization''~\cite{bakunov2000adiabatic, Ilya2010_1, Ilya2010_2, Kenan_2018_upshift}, with which the plasma is ionized at an infinitely slow rate. These two extremes predict different output amplitudes. 

However, the ionization time of laboratory plasmas generally falls into the intermediate regime between a few laser cycles and a few nanoseconds~\cite{mulser2010high, Starace_PRL2002, Dodin_PRE2003}. Thus, a theory addressing experimentally applicable parameter regimes is lacking for designing experiments and verifying the results. The reported experiments~\cite{KuoExp1993,kuo1990frequency, yugami2002experimental, Suckewer2002, Suckewer2005, Nishida2012, Shvets2017} lack detailed comparison of energy conversion efficiencies to a first-principles theory.  Modeling the effect of ionization speed is particularly important in recent development of ``flying focus''~\cite{Froula2018, Turnbull_2018, Howard2018, Howard2019}, which allows continued frequency upconversion by solving the issues of group velocity mismatch and pump pulse diffraction.

To provide the background for analyzing the effects of finite ionization rates, we first briefly review the process of ``flash ionization''~\cite{wilks1988frequency, Kenan_2018_upshift}. All the electrons are suddenly ionized and begin to oscillate in the laser field. The electron oscillations constitute the polarization current~\cite{Kenan_2018_upshift}, which reduces the index of refraction and causes frequency upconversion. The electron oscillation also radiates waves in both forward and backward directions, thereby splitting the input pulse into a pair of counter-propagating pulses. Nevertheless, the summation of the electric field amplitudes of the pulse pair does not change~\cite{Kenan_2018_upshift}. The electron oscillation motion and the laser field has a $\pi/2$ phase shift. Since the initial electron momentum is zero, the electric fields that are not at the maximum-amplitude phase induce a static transverse electron drift, whose amplitude is proportional to the total electric field amplitude. Therefore, ionization partitions the input pulse energy into a pair of counter-propagating pulses and transverse electron drift with zero frequency. 

A subsequent ionization process redistributes the energies of the counter-propagating pulse pair and increases the transverse current energy. If the input laser field is a continuous plane wave, the amplitudes of the forward and backward propagating pulses have a certain ratio at a given frequency, because they are both proportional to the amplitude of the polarization current. The total energy carried by the transverse electron drift would also be of the same proportion to the total electric field amplitude, which is an invariant during ionization. Therefore, the energy partition of ``two-step'' ionization is identical to the result of ``flash ionization'' when achieving the same frequency upconversion.  The same argument can be extended to continuously ionized plasmas to conclude that finite ionization rate does not change the energy conversion efficiency of a continuous plane-wave input. 
With slow ionization, only the transverse current has a smaller amplitude than ``flash ionization'' because the electrons can drift in the opposite directions. 

However, the energy partition would change if the input laser has finite duration. After splitting into two  pulses, the two pulses emanating from the ionization region propagate in different directions and do not share the same polarization current. As a result, the forward- and backward-propagating wave amplitudes no longer have the same proportionality relation. The pulse separation also reduces the summation of electric field amplitudes, leading to a different amplitude of transverse current. If the pulse is sufficiently short, or if the ionization rate is sufficiently low, such that the backward-propagation wave can be neglected, the analysis reduces to the ``adiabatically gradual ionization'' limit~\cite{bakunov2000adiabatic, Ilya2010_1, Ilya2010_2, Kenan_2018_upshift}. However, for pulses with finite duration and for finite ionization rates, the counter-propagation pulses could partially overlap during ionization. The energy partition becomes position dependent, and the energy conversion efficiency into the transmitted pulse is affected by the overlapping volume. Describing the pulses evolution requires a set of coupled wave equations.

Here, we describe the laser amplitude dynamics when the mediating plasmas have different, finite ionization rates. We derive a set of coupled wave equations for laser propagation in plasmas ionized by an external field, including one created with a ``flying focus''. For homogeneous plasmas, we analytically solve the equations for the energy conversion efficiencies with different pulse durations and ionization rates. We find that the amplitude of the frequency-upconverted pulse is determined by the product of the pulse duration and rate of frequency change, \ie the total frequency change within the pulse duration. We further find the conditions of achieving the maximum transmission energy, and achieving the maximum reflection energy.

The paper is organized as follows: In Sec.~\ref{model}, we describe the electron motion when the plasma is ionized and we derive coupled wave equations for the laser field amplitudes at finite ionization rates. In Sec.~\ref{analysis}, we then consider evolution of pulses with different durations and ionization rates. We find the conditions for maximum transmission and for maximum reflection. In Sec.~\ref{expcomp}, we compare our results with published experiment data. In Sec.~\ref{concl}, we discuss our results.

\section{Model} \label{model}

Consider an electromagnetic wave described by electric field $\bm{E}$ and magnetic field $\bm{B}$ in a gas medium. Suppose that the gas molecules are then ionized by a different field. Propagation of the electromagnetic wave in the plasma is then described by the wave equations
\begin{align}
	\nabla^2\bm{E} -\frac{1}{c^2} \frac{\partial^2}{\partial t^2} \bm{E} &= \frac{1}{c^2\varepsilon_0} \frac{\partial}{\partial t} \bm{J},  \label{1} \\
	\nabla^2\bm{B} -\frac{1}{c^2} \frac{\partial^2}{\partial t^2} \bm{B} &= -\mu_0 \nabla \times \bm{J},  \label{2}
\end{align}
where $\varepsilon_0$ and $\mu_0$ are the permittivity and permeability, respectively, of vacuum and $\varepsilon_0\mu_0 = 1/c^2$. The equations (\ref{1}) and (\ref{2}) show that the fields $\bm{E}$ and $\bm{B}$ depend on the current $\bm{J}$. 

In an ionizing plasma, the current is carried by electrons that are oscillating in the electromagnetic field $\bm{J} {=-} e \sum_i n_i \bm{v}_i  $, where $n_i $ is the density of the electrons created at a certain time $t_i$ and $\bm{v}_i $ is the electron drift velocity at ionization. We distinguish the electrons ionized at different times, because, while each of the electron groups can be described by the fluid model, they together cannot be a single fluid. Note that, although the external ionization field or the collision of particles induce nonzero initial momentum to the ionized electrons especially in case of above-threshold ionization~\cite{Starace_PRL2002, Dodin_PRE2003}, the specific type of electron drift motion caused by the photon or phonon momentum has zero frequency. Hence, they do not coherently interact with the electromagnetic wave, and can be neglected in our model. 
What we are interested in is the polarization current which is carried by the electron oscillation at the same frequency of the electromagnetic wave. This type of electron oscillation motion is driven by the electric field $\dot{\bm{v}}_i  {=} -(e/m_e)\bm{E}$. We can neglect the influence of the magnetic fields $\bm{B}$ in the non-relativistic regime. We denote the total electron density as $n=\sum n_i $, and then obtain 
\begin{equation} \label{3} 
\frac{1}{\varepsilon_0}\frac{\partial}{\partial t} \bm{J} = \frac{e^2}{m_e\varepsilon_0} n \bm{E} \equiv \omega_p^2 \bm{E} ,
\end{equation} 
where $\omega_p=\sqrt{e^2n/(m_e\varepsilon_0)}$ is the plasma frequency. 
Integrating, we can write the current as 
\begin{align} 
\bm{J} &= \frac{e^2n}{m_e} \int_{-\infty}^t \bm{E}(t') \md t' + \bm{J}_s,  \label{4a} \\
\bm{J}_s &= - \int_{-\infty}^t \frac{e^2}{m_e} \sum_i\dot{n}_i (t') \int_{-\infty}^{t'} \bm{E}(t'') \md t'' \md t'  \label{4b} 
\end{align}
where $\dot{n}_i $ denotes the incremental density due to ionization of the electrons at the time $t_i$. In Eq.~(\ref{4a}), the first term on the right hand side oscillates at the electromagnetic frequency and creates the polarization current. It is responsible for the decrease of optical refractive index and for the increase of wave frequency~\cite{Kenan_2018_upshift}.  
The $\bm{J}_s$ term describes a static transverse current resulting from the phase mismatch between the oscillating electromagnetic field and the ionized electrons. The value of $\bm{J}_s$ depends on the final plasma density and, more importantly, the time of ionization. For ``flash ionization'', all the electrons drift with the same phase, so the average value of $\bm{J}_s$ is finite. In theory, $\bm{J}_s$ could be zero (or a maximum value)  if all the electrons are ionized when the instantaneous $\bm{E}$ field is at its peak (trough) value.  However, the synchronization of ionization and the wave phase is impractical. For finite ionization rates, static currents are generated with distributed phases of $\bm{E}$ at different times. The electrons drift in all the directions, and hence the average static current becomes negligible. In this case, the energy eventually dissipates as heat.


The equations for the electromagnetic wave dynamics can be obtained by substituting Eqs.~(\ref{3})-(\ref{4b}) into Eqs.~(\ref{1})-(\ref{2})
\begin{align}
	c^2 \nabla^2\bm{E} -\frac{\partial^2}{\partial t^2} \bm{E} &= \omega_p^2 \bm{E} ,  \label{5} \\
	c^2 \nabla^2\bm{B} -\frac{\partial^2}{\partial t^2} \bm{B} &= \omega_p^2 \bm{B} - \int_{-\infty}^t \frac{\partial \omega_p^2}{\partial t'}  \bm{B} \md t' .  \label{6}
\end{align}
Here, the expressions are simplified using Faraday's law and the relation $\partial\omega_p^2/\partial t = e^2 \dot{n}/(m_e\varepsilon_0)$. 

At a certain time $t$, the solution of Eqs.~(\ref{5}) and (\ref{6}) describes a pair of counter-propagating waves at the same frequency and a static magnetic field
\begin{align} 
\bm{E} &= \left[ E_+ e^{-i(\omega t-\bm{k}\cdot\bm{r})} + E_- e^{i(\omega t + \bm{k}\cdot\bm{r})} \right] \hat{e} , \label{7} \\
\bm{B} &= \left[ B_+ e^{-i(\omega t-\bm{k}\cdot\bm{r})} + B_- e^{i(\omega t + \bm{k}\cdot\bm{r})} + B_s e^{i\bm{k}\cdot\bm{r}} \right] \hat{e} , \label{7b}
\end{align}
where $\hat{e}$ is the direction of polarization, $\omega$ and $k$ are the wave frequency and wavevector, respectively, obeying the dispersion relation $\omega^2 {=} \omega_p^2 {+} c^2k^2$. $E_\pm$ and $B_\pm$ are the amplitude of the forward-/backward-propagating electric and magnetic fields, respectively, and they are related to the electric field amplitude by $B_\pm = \pm(ck/\omega) E_\pm$. $B_s$ denotes the amplitude of the static magnetic field generated by the static current $\bm{J}_s$. 


We next use the slowy varying envelope approximation to reduce Eqs.~(\ref{5}) and (\ref{6}) into first-order differential equations. To evaluate the integral, we use the relations $\partial \omega_p^2/\partial t = \partial \omega^2/\partial t = 2\omega \dot{\omega}$ and $\dot{\omega}/\omega \ll \omega$. Then, we obtain a set of equations of the forward and backward propagating modes 
\begin{align}
	\omega \frac{\partial}{\partial t} (E_+ - E_-) - c^2 k \nabla (E_+ + E_-) &= 0 , \label{8}\\
	\omega \frac{\partial}{\partial t} (B_+ - B_-) - c^2 k \nabla (B_+ + B_-) &= -\dot{\omega}(B_+ - B_-). \label{9}
\end{align}
The equations can be reduced to a pair of coupled wave propagation equations by using the relation $B_\pm = \pm(ck/\omega) E_\pm$ as 
\begin{align}
	 \frac{\partial}{\partial t} E_+ - v_g  \nabla E_+ + \frac{\dot{\omega}}{2\omega} E_+ &= \frac{\dot{\omega}}{2\omega} E_- , \label{10}\\
	 \frac{\partial}{\partial t} E_- + v_g  \nabla E_- + \frac{\dot{\omega}}{2\omega} E_- &=  \frac{\dot{\omega}}{2\omega} E_+ , \label{11}
\end{align}
where $v_g = c^2k/\omega$ is the group velocity of the wave. The factor $\dot{\omega}$ can be both space and time dependent, which determines the region of interaction. 
The coupled wave equations (\ref{10}) and (\ref{11}) can describe the electromagnetic wave propagation in an arbitrarily ionized plasma specified by the parameters $\dot{\omega}(t, \bm{r})$ and $\omega(t, \bm{r})$. The solution can be found numerically if the profile of plasma ionization is complicated. 

\begin{figure}[thp]
	\includegraphics[width=0.8\linewidth]{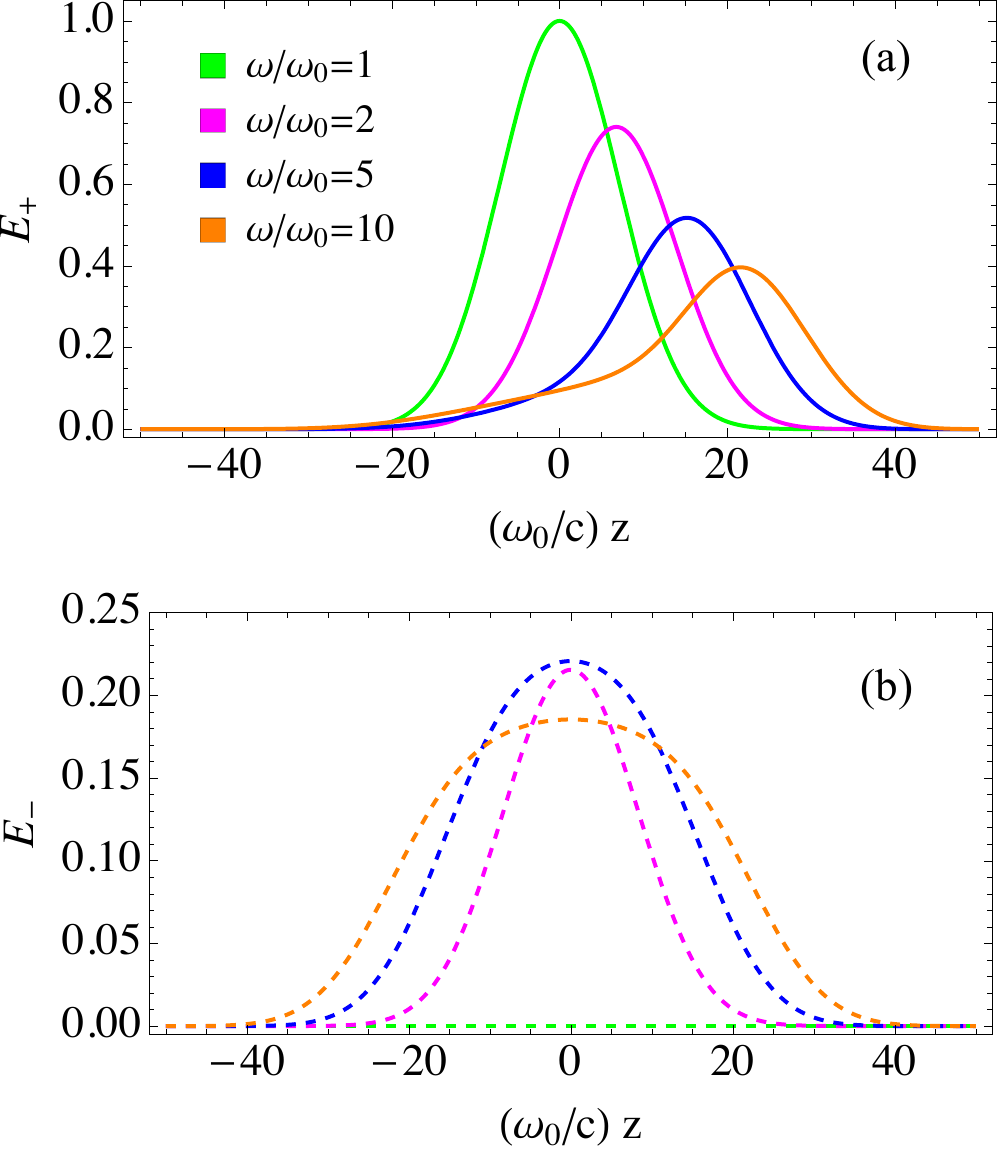}
	\caption{ Time evolution of the transmitted pulse (top panel) and reflected pulse (bottom panel) when the pulse frequency is upconverted from $\omega_0$ to $10\omega_0$ in an ionizing plasma. The laser frequency increases as $\omega(t) = \omega_0 + \dot{\omega} t$ where $\dot{\omega}=0.1\omega_0^2$. The initial pulse envelope is $e^{-[z/(\omega_0 c \tau)]^2}$ with $\omega_0\tau=5$. } 
	\label{evo}
\end{figure}

Equations.~(\ref{10}) and (\ref{11}) show that both the counterpropagating modes ($E_+$ and $E_-$) damp and couple to each other at the same rate $\dot{\omega}/(2\omega)$. However, more energy flows from the $E_+$ mode to the $E_-$ mode because $E_+$ has a larger amplitude than $E_-$. The consequence of the non-reciprocal energy flow is two fold. First, it leads to an energy loss of the $E_+$ mode and an energy gain of the $E_+$ mode. Second, the total energy of the $E_+$ and $E_-$ modes decreases. The energy is lost in the form of static electron currents which induce either a static magnetic field in fast ionization or heat in slow ionization.


\begin{figure*}[bht]
	\includegraphics[width=\linewidth]{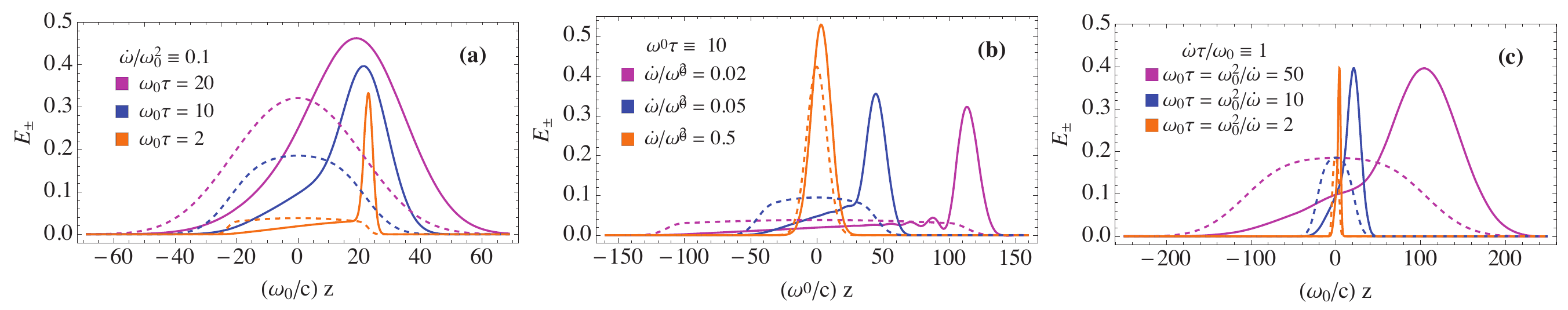}
	\caption{ Pulse envelopes of the the forward (solid curves) and backward (dashed curves) propagation modes after the pulse frequency is upconverted from $\omega_0$ to $10\omega_0$. The three pairs of curves in each plot compare (a) different ionization rates $\dot{\omega}$ but fixed pulse duration $\tau$, (b) different $\tau$ but fixed $\dot{\omega}$, and (c) different  $\dot{\omega}$ and $\tau$ but fixed $\dot{\omega}\tau$.  } 
	\label{comp}
\end{figure*}

To illustrate the coupling between the forward- and backward-propagation modes during frequency upconversion, we show, in Fig.~\ref{evo}, an example of the envelope evolution of a laser pulse in an ionizing plasma by solving the coupled wave equations~(\ref{10}) and (\ref{11}). The initial laser pulse (green solid curve) has a Gaussian-shape envelope $E_0 = e^{-(z/c\tau)^2}$ with ${\omega_0\tau=5}$. The laser frequency increases linearly $\omega(t) = \omega_0 + \dot{\omega} t$ where $\dot{\omega}=0.1\omega_0^2$. The magenta, blue, and orange curves show the pulse envelopes of the forward- and backward-propagating modes $E_\pm$ when the pulse frequency is subsequently upconverted to $2\omega_0$, $5\omega_0$, and $10\omega_0$, respectively. 
We observe that, when the frequency upconversion begins, $E_+$ quickly loses energy. The lost energy is partially transformed into the $E_-$ mode, which forms an envelope that resembles the shape of the initial pulse $E_0$. As the pulse frequency $\omega$ increases, the $E_+$ mode continues to lose energy, but at a slower rate due to the reduction of the interaction strength $\dot{\omega}/(2\omega)$. The pulse propagation also slows down for a reduced group velocity $v_g\propto \omega^{-1}$. The $E_+$ mode develops a long tail and the $E_-$ mode transforms from a Gaussian shape into a flat-top-pulse shape. While the peak amplitude of $E_+$ decreases monotonically, the peak amplitude of $E_-$ first increases and then begins to decrease at the point when ${E_+(z=0)} \leq {E_-(z=0)}$. The $E_-$ mode continuously couples to the tail of $E_+$, leading to an asymmetric shape of $E_+$.

\section{Effects of finite ionization rate and finite pulse duration} \label{analysis}

The pulse amplitudes, after frequency upconversion, are determined by the interaction between the counter-propagating modes. At a certain ionization rate, a longer pulse duration allows a longer growth time leading to a larger amplitude of the $E_-$ mode. On the other hand, a stronger $E_-$ also translates more energy into the $E_+$ and maintaining a larger amplitude of $E_+$ after frequency upconversion.

To illustrate the effects of finite pulse duration and ionization rate, we compare the pulse envelopes when different pulses are frequency-upconverted from $\omega_0$ to $10\omega_0$ and show the $E_+$ and $E_-$ modes as solid and dashed curves in Fig.~\ref{comp}. Figure.~\ref{comp}(a) compares different pulse durations in the same ionizing plasma. The curves show that a longer pulse can achieve a larger final amplitude due to longer interaction time between the $E_+$ and $E_-$ modes. The longer interaction time also causes a shift of the peak towards the $-z$ direction at larger $\tau$. The decrease of pulse amplitude and a change in the $E_-$ envelope shape can also be seen when the ionization rate $\dot{\omega}$ decreases, which can be observed in Fig.~\ref{comp}(b). 

Actually, the independent parameter that determines the pulse peak amplitude becomes apparent when rewriting Eqs.~(\ref{10}) and (\ref{11}) as 
\begin{align}
	\frac{\omega}{\dot{\omega}} \frac{\partial}{\partial t} E_+ - \frac{c^2k}{\dot{\omega}}  \frac{\partial}{\partial z} E_+ &= - \frac12(E_+ - E_-) , \label{12}\\
	\frac{\omega}{\dot{\omega}} \frac{\partial}{\partial t} E_- + \frac{c^2k}{\dot{\omega}}  \frac{\partial}{\partial z} E_- &=  \frac12(E_+ - E_-). \label{13}
\end{align}
The form of the derivative terms indicates that $\dot{\omega}$ yields a contraction of time and distance such that $t\to \dot{\omega}t/\omega$ and $z \to \dot{\omega}z/(c^2k)$. Evolution of the pulse amplitudes $E_\pm$ is determined by the normalized pulse duration $\dot{\omega}\tau/\omega_0^2$ and the total change of frequency. 
The simulation results of pulse peaks with a fixed normalized pulse duration $\dot{\omega}\tau/\omega_0^2=1$ are shown in Fig.~\ref{comp}(c). The curves show that the different pulses achieve the same amplitudes despite different ionization rates. 
The effect of a slow ionization rate and a long pulse duration only stretches the reflection mode $E_-$ into a flat-top shape when the ionization rate decreases.  This change in the reflection pulse shape might affect certain applications. For example, shorter pulse duration but larger ionization rate are favored in order to obtain sharper frequency-upconverted reflections.

\begin{figure}[tp]
	\centering
	\includegraphics[width=\linewidth,valign=t]{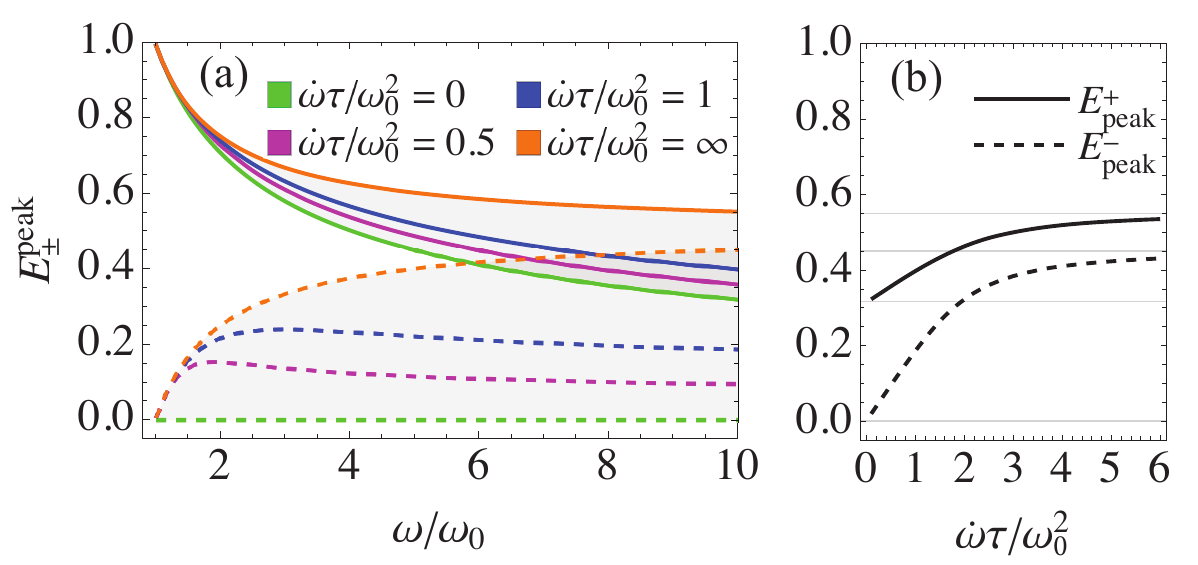}
	\caption{ (a) Evolution of the pulse peak amplitudes when its frequency is upconverted from $\omega_0$ to $10\omega_0$ in an ionizing plasma. The solid curves and dashed curves show the forward and backward propagation modes. The four pairs of curves compare different pulse duration ranging from $\omega_0 \tau=0$ (limit of $\delta$-function shape) to $\infty$ (limit of continuous wave). (b) The peak amplitude of $E_\pm$ when the pulse frequency reaches $10\omega_0$. The thin grid lines show the asymptotic values of $E_\pm$ in the long/short pulse limit for $\omega=10\omega_0$. } 
	\label{coe}
\end{figure}

The parameters of normalized time $\dot{\omega}t/\omega$ and normalized space $\dot{\omega}z/(c^2k)$ in Eqs.~(\ref{12}) and (\ref{13}) can provide an equivalence condition between high ionization rate and long pulse duration on the energy conversion efficiency. It is important in experiments because longer pulses can be used to substitute the need for faster ionization rates. 
In the asymptotic limit of infinite pulse duration $\dot{\omega}\tau/\omega_0^2 \to \infty$, the value of $E_\pm$ becomes spatially invariant. The solution to Eqs.~(\ref{12}) and (\ref{13}) can be easily obtained by neglecting the spatial derivatives terms 
\begin{equation}\label{21}
E_{\pm} = \frac12 \left( 1 \pm \frac{\omega_0}{\omega}\right) E_0,
\end{equation}
where $\omega_0$ and $\omega$ denote the initial and final pulse frequency. The evolution of $E_{\pm}$ duration frequency upconversion is displayed as the orange curves in Fig.~\ref{coe}. This expression (\ref{21}) is identical to the ones reported in Refs.~\cite{wilks1988frequency, Kenan_2018_upshift}. However, our derivation shows that the energy conversion efficiency of a continuous wave is independent of the ionization rate. The equal wave energy partition is achieved at the high frequency limit regardless of the speed of ionization. The universality of formula (\ref{21}) is the result of reciprocal energy transitioning between two continuous waves: As the $E_+$ mode loses energy to the $E_-$ mode when plasma is ionized, the $E_-$ mode energy is also converted to the $E_+$ mode. The dynamic balance is achieved when $E_+ =E_-$, \ie the forward and backward propagating modes have the same amplitude. 

In the limit of $\dot{\omega}\tau/\omega_0^2 \to 0$, the amplitude of $E_-$ mode becomes negligible. The evolution of the pulse amplitude can be obtained by solving Eq.~(\ref{10}) and neglecting the $E_-$ terms, and 
\begin{equation}\label{31}
E_{+} = \sqrt{\frac{\omega_0}{\omega}} E_0. 
\end{equation}
The pulse peak amplitude is plotted as the solid green curve in Fig.~\ref{coe}, which shows that $E_+$ varies inversely with the square root of the pulse frequency. The relation has an identical form with energy conversion efficiency with infinitesimal ionization rates~\cite{bakunov2000adiabatic, Ilya2010_1, Ilya2010_2, Kenan_2018_upshift}. But our result shows that the same relation applies to $\delta$-function-shape pulses even when the ionization rate is finite.

For finite values of pulse durations and ionization rates, the pulse evolution depends on the specific pulse envelope and it is difficult to find an analytical solution in general. We numerically solve Eqs.~(\ref{10}) and (\ref{11}) with the parameters $\dot{\omega}\tau/\omega_0^2 {=0.5}$ (magenta) and ${=1}$ (blue). The evolution of the peak amplitudes are displayed in Fig.~\ref{coe}(a), and the output peak amplitudes are displayed in Fig.~\ref{coe}(b). Since the electromagnetic wave energy density in plasma is 
\begin{equation}
W_\pm=\varepsilon_0 E_\pm^2, 
\end{equation}
the curves also describe the energy density distribution of the two modes. The curves reveal that the peak value $E_+^\mathrm{peak}$ decreases monotonically as the frequency increases. The decrease is slower with longer pulse duration $\dot{\omega}\tau/\omega_0^2$ and the output is stronger when the pulse frequency reaches $10\omega_0$. The solid curve in Fig.~\ref{coe}b shows that the output amplitude approaches the asymptotic value when the pulse duration is much longer than the ionization time. Figure~\ref{coe}a also shows that the peak amplitude of the reflection mode $E_-^\mathrm{peak}$ first increases and then decreases for finite pulse durations. The maximum value of $E_-^\mathrm{peak}$ at near $\omega/\omega_0\sim 1$ corresponds to the time when the input pulse propagates for a distance of $\tau$, beyond which the $E_-$ mode begins to transfer energy back to the tail of the $E_+$ mode. 
Hence, the strongest output pulses can be obtained when the plasma is fully ionized in a time frame that is shorter than the pulse duration.

\section{Comparison with experiment} \label{expcomp}

\begin{figure}[tp]
	\centering
	\includegraphics[width=0.8\linewidth,valign=t]{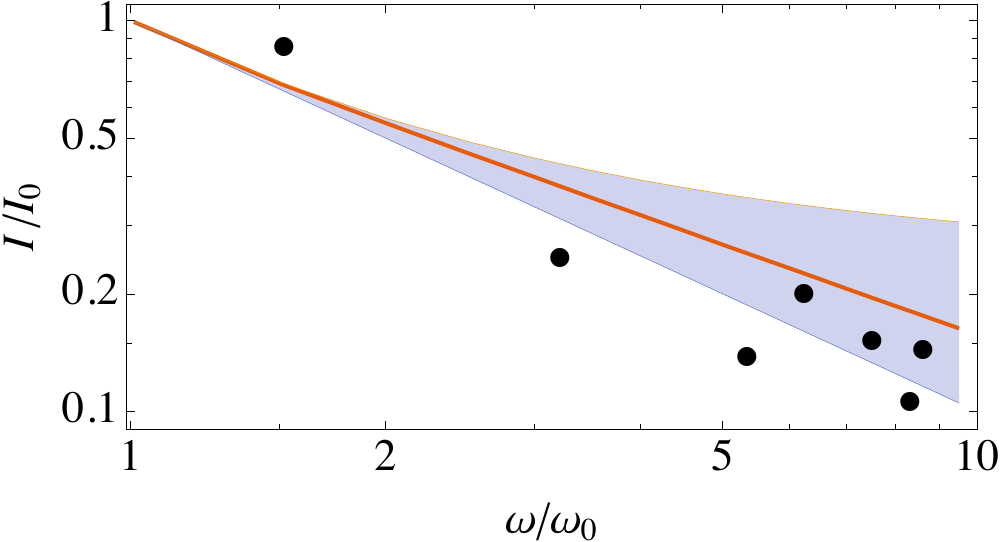}
	\caption{ Comparison between the theory and experiment values of the transmitted pulse intensity at different upconverted frequencies. The shaded area shows the pulse intensity between sudden ionization (top) and gradual ionization (bottom). The solid curve shows the pulse intensity with an intermediate ionization time that equals to the pulse duration. The dots are the experiment data retrieved from Ref.~\cite{Nishida2012}, with the permission of AIP Publishing. } 
	\label{exp}
\end{figure}

For verification, we compare our theory with the data published by Nishida, \etal~\cite{Nishida2012}. The experiment~\cite{Nishida2012} achieves up to ten-fold frequency upconversion of $0.35\, \mathrm{THz}$ pulses by ionizing a ZnSe crystal. Variation of the upconverted frequencies are achieved by controlling the plasma density using different ionization laser intensities. The experiment demonstrates dependence of the output pulse intensity on the upconverted frequencies, which is plotted as black dots in Fig.~\ref{exp}. Since the ZnSe crystal has a $2\, \mathrm{mm}$ thickness and a $68^\circ$ tilt angle, it can only carry less than $15$ cycles of the terahertz pulses. 
Ionization time is not given, but estimation based on a similar experiment~\cite{Suckewer2005} suggest that at least ten pulse cycles have elapsed by the time the ionization pulse is at its maximum intensity $1.4{\times} 10^9\, \mathrm{W/cm}^2$, corresponding to the maximum frequency upconversion. When weaker ionization pulses are used for smaller frequency upconversion, the ionization time could be even longer. Thus, the ionization in the experiment is close to the gradual ionization limit when $\omega/\omega_0$ is small and the ionization time is similar to the pulse duration when $\omega/\omega_0$ is large. 

The experiment data are compared with the theory values in Fig.~\ref{exp}. The solid curve shows the output pulse intensity if the ionization time is the same with the pulse duration.  The shaded region is enclosed by the sudden ionization limit (top) and the gradual ionization limit (bottom). Since the pulse energy is partially converted and the signal collection efficiency is non-perfect, the true intensity of the frequency upconverted pulse could deviate from the displayed data points. Nevertheless, the dots in Fig.~\ref{exp} reveal a trend which is closer to the gradual ionization limit when $\omega/\omega_0$ is small, and approaches the solid curve when $\omega/\omega_0$ is large. Thus, our theory shows a decent agreement with the experiment data.

\section{Conclusion} \label{concl}

In conclusion, we derive a set of coupled differential equations to describe the evolution of a laser pulse when its frequency is upconverted in an ionizing plasma. We find that the energy conversion efficiency can be optimized by controlling both pulse duration as well as plasma ionization rate.  It complements the existing theories~\cite{morgenthaler1958velocity, wilks1988frequency, mendoncca2000book, nerukh2012non, kalluri2016electromagnetics, wilks1988frequency, bakunov2000adiabatic, Ilya2010_1, Ilya2010_2, Kenan_2018_upshift} which solely describe the effects of different ionization rates. The strongest transmission and reflection can be obtained when the plasma is fully ionized in a time that is shorter than the pulse duration, which is evidenced by experiment data. The same rule for the maximum intensity also applies to frequency upconversion with a ``flying focus''~\cite{Froula2018, Turnbull_2018, Howard2018, Howard2019}. 

The underlying physics of the maximum energy conversion efficiency is apparent by analyzing the polarization current, which is responsible for the frequency upconversion. The polarization current partially reflects the input wave energy into a frequency-upconverted backward-propagating mode. Since the coupling strength is proportional to the difference of the two counter-propagating modes, spatial overlapping of the modes reduces the energy outflow of the input wave. 
The mismatch of the phase of electric field and electron initial momentum also leads to non-polarization static currents. However, the currents are generated with different phases and average to zero when the plasma is ionized in multiple laser periods. The energy of non-polarization current then dissipates as heat.

The coupled differential equations~(\ref{10}) and (\ref{11}) are both space and time dependent and hence they can describe more complex scenarios involving spatially inhomogeneous and temporally evolving plasma densities, \eg the evolution of the pulse envelope and frequency of an pulse inside a ``flying focus''. They can also describe backward Raman amplification with an ionizing front~\cite{Malkin_pop2001, Clark_pop2002, Clark_pop2003, Zhang_pop2014}. When implementing the equations in a numerical code, it should be born in mind that the group velocity is frequency-dependent. The decrease of group velocity becomes significant especially when the plasma frequency is much higher than the initial pulse frequency, \ie the wave frequency and plasma frequency are similar. The decreased group velocity causes longer overlapping time between the counter-propagating mode and hence can increase the energy conversion efficiency.

\begin{acknowledgments}
	This work was supported by NNSA Grant No. DE-NA0002948, and AFOSR Grant No. FA9550-15-1-0391.
\end{acknowledgments}  

\bibliography{slowupshift}

\end{document}